# Quasi-Perpendicular High Mach Number Shocks

A. H. Sulaiman,[1] A. Masters,[1] M. K. Dougherty,[1] D. Burgess,[2] M. Fujimoto,[3] G. B. Hospodarsky[4]

[1]Space and Atmospheric Physics, Blackett Laboratory, Imperial College London, London, UK.

[2]Astronomy Unit, Queen Mary University of London, London, UK.

[3]Institute of Space and Astronomical Science, Japan Aerospace Exploration Agency, 3-1-1 Yoshinodai, Chuo-ku, Sagamihara, Kanagawa 252-5210, Japan.

[4]Department of Physics and Astronomy, University of Iowa, Iowa City, Iowa, USA.

Corresponding author: A. H. Sulaiman

Corresponding author email: ali.sulaiman08@imperial.ac.uk

Abstract

Shock waves exist throughout the universe and are fundamental to understanding the nature of collisionless plasmas. Reformation is a process, driven by microphysics, which typically occurs at high Mach number supercritical shocks. While ongoing studies have investigated this process extensively both theoretically and via simulations, their observations remain few and far between. In this letter we present a study of very high Mach number shocks in a parameter space that has been poorly explored and we identify reformation using *in situ* magnetic field observations from the Cassini spacecraft at 10 AU. This has given us an insight into quasi-perpendicular shocks across two orders of magnitude in Alfvén Mach number ($M_A$) which could potentially bridge the gap between modest terrestrial shocks and more exotic astrophysical shocks. For the first time, we show evidence for cyclic reformation controlled by specular ion reflection occurring at the predicted timescale of ~0.3 $\tau_c$, where $\tau_c$ is the ion gyroperiod. In addition, we experimentally reveal the relationship between reformation and $M_A$ and focus on the magnetic structure of such shocks to further show that for the same $M_A$, a reforming shock exhibits stronger magnetic field amplification than a shock that is not reforming.

Background

The solar wind flow is both supersonic and super-Alfvénic, so that as it interacts with a magnetized planet such as Saturn, a bow shock wave forms in the flow upstream of the magnetosphere. The planetary bow shocks, from Mercury to Uranus provide a unique resource in terms of parameter range for understanding shocks in other astrophysical systems. At the bow shock the flow is slowed and heated, and the density and magnetic field increase, with conservation of mass, momentum and energy giving a set of jump conditions which locally relate the flow parameters upstream and downstream of the shock. The solar wind is essentially collisionless, with a collisional mean free path orders of magnitude greater than the width of the shock transition layer. Therefore the transition is attained by coupling between the electric and magnetic fields and the particles, involving both microinstabilities and particle trajectories in the macroscopic fields of the shock transition. The interactions depend in complex ways of the shock parameters such as the Mach number $M$, the ratio of the relative flow speed to a characteristic wave speed, the upstream plasma beta $\beta$, the ratio of thermal to magnetic pressures, and $\theta_{Bn}$, the angle between the upstream magnetic field vector $\boldsymbol{B}$ and the normal to the shock front $\hat{\boldsymbol{n}}$. Due to their conducting and magnetized nature, space plasmas can support several waves, such as non-compressional Alfvén waves and compressional sound and magnetosonic waves. Consequently, a shock may be characterized by different Mach numbers, such as the Alfvén Mach number $M_A$, the sonic Mach number $M_s$, and the fast magnetosnonic Mach number $M_f$.

Within the framework of a fluid description where dissipation is provided by resistivity (or in the case of a collisionless plasma, by anomalous resistivity), shock solutions are possible up to a limiting Mach number, the critical Mach number $M_c$. The critical Mach number corresponds to when the flow speed immediately downstream equals the sound speed, i.e. the sonic Mach number of the downstream flow is unity ($M_{s,d} = 1$), and it is found that $1 \leq M_c \leq 2.76$, depending on the shock parameters [1]. For supercritical shocks ($M > M_c$), there is excess energy in the directed bulk flow that cannot be converted into thermal energy within the current-carrying transition layer in the timescale of the fluid element's crossing since the energy dissipation required by the jump conditions cannot be provided solely by resistive heating. Observations and simulations show that supercritical shocks compensate for the shortfall in dissipation by reflecting some fraction of the incoming ions back upstream [2]. At such high Mach number shocks, the structure becomes inherently dependent on the ion dynamics, and a fluid description is inadequate. Saturn's orbit is at a heliocentric distance of ~10 AU (1 AU = $1.5 \times 10^8$ km), a region characterized by significantly higher Mach numbers normally not accessible in near-Earth space. The Cassini spacecraft therefore offers a unique opportunity to investigate the near-Saturn plasma conditions in addition to its principal objectives in planetary science.

A quasi-perpendicular shock is defined as having an angle $\theta_{Bn} > 45°$. Figure 1 illustrates *in situ* magnetic field signatures for a typical outbound (i.e. passing from downstream to upstream) crossing of a quasi-perpendicular shock for $\theta_{Bn} = 70°$. The crossing is characterized by a sharp, local transition between both regimes in contrast with a quasi-parallel shock ($\theta_{Bn} < 45°$) where the transition layer extends from a foreshock region far upstream, excited by wave-particle interactions, to the downstream region [3,4]. The ramp of a quasi-perpendicular shock is immediately preceded by a foot which spans a distance comparable to the gyroradius of the reflected ions. The locality of the foot corresponds to the spatial restriction imposed on the specularly reflected ions by the orientation of the magnetic field being close to parallel to the shock front. These ions respond to the transverse convective electric field $-\boldsymbol{V} \times \boldsymbol{B}$ from which they gain sufficient energy during the course of their gyration around the magnetic field until they return to the shock and are eventually transmitted across [5].

Although ion reflection dominates heating at high Mach number shocks, in recent years, attention has focused on variability or nonstationarity in the shock structure at ion time scales. There are several proposed mechanisms, mostly based on simulations and theoretical considerations [6,7]. At sufficiently high Mach number and low upstream ion beta $\beta_i$, there is a quasi-periodic, cyclic reformation of the shock associated with over-reflection of ions [8,9,10]. This mechanism predicts a time scale for reformation of the order of the ion cyclotron period $\Omega_c^{-1}$ [11]. Nonstationarity has also been suggested to be the outcome of a gradient catastrophe of nonlinear upstream whistler, associated with Mach numbers greater than the (nonlinear) whistler critical Mach number beyond which an upstream whistler cannot phase stand in the upstream flow [12]. An alternative mechanism found in particle-in-cell (PIC) simulations is the quasi-periodic disruption of the ion foot due to the modified two-stream instability [13].

Shock reformation has primarily been studied using hybrid and PIC simulation, where a comprehensive picture of the time evolution of the shock structure is available. *In situ* spacecraft observations have neither been frequent [14] nor extensive enough to corroborate these studies. This is mainly due to most studies of shock crossings being near the Earth where the Mach numbers typically range from low to modest ($M_A$ = 2-8). While some observations have been reported at Earth's bow shock [15], they remain open to interpretation [16]. In this letter, we present observations of very high Mach number shock reformation using data from the Cassini spacecraft's fluxgate magnetometer (MAG) [17]. We construct a parameter space of Alfvén Mach number $M_A$ that spans two orders of magnitude focussing particularly on the underexplored highest $M_A$ regime. At such heliocentric distances the regularly azimuthal interplanetary magnetic field (IMF), makes a quasi-perpendicular shock and this is the most probable configuration at any given encounter.

Method

We analyse data chiefly from the magnetometer and use the plasma instruments, the Radio and Plasma Wave Science (RPWS) [18] and the Ion Mass Spectrometer (CAPS-IMS) [19], where available. The limited plasma datasets have not undermined in-depth studies of the physical properties of the bow shock and neighbouring regions [20,21,22]. The magnetic field data is at 1s time resolution.

We accumulated shock crossings spanning from 2007 to 2012 and calculate $M_A$ from the equation below

$$M_A \equiv \frac{u}{v_A} = \frac{\sqrt{P_{dyn}}}{B_u}\sqrt{\mu_0} \qquad (1)$$

where $P_{dyn}$ is the upstream ram pressure $\rho u^2$ and $B_u$ is the upstream magnetic field strength. Given that the three-dimensional shape and size of Saturn's bow shock are known [23], we are able to determine the subsolar distance $R_{SN}$ of the shock from any point it is crossed. We are then able to work out $P_{dyn}$ using the power law, $R_{SN} \propto P_{dyn}^{-1/5.4}$, which was derived from an empirical fit to many crossings. Embedded in this relationship are local density measurements and solar wind propagations. By obtaining $P_{dyn}$ and measuring $B_u$ for each crossing, we can then calculate $M_A$ using equation (1). For this work, we have validated this method of obtaining $M_A$ with ion densities and solar wind speeds (in the shock's rest frame) from the plasma instruments where both are available and unambiguous. $M_A$ is calculated using the local normal component of the flow velocity, where the local normal $\hat{n}$ is obtained from a modelled shape of the shock surface [23]. This method has been shown to be substantially more accurate, using a single spacecraft, than the coplanarity theorem [24]. Table 1 lists all the relevant parameters.

Results and Conclusion

Figure 2 (a-c) show three examples of quasi-perpendicular crossings from upstream of the shock to downstream (right to left) using magnetic field data. These are three examples revealing the foot signatures, of enhanced magnetic field strength upstream, occurring at regular intervals suggestive of reformation cycles (with the frequency also present in the downstream magnetic field). These pulses are attributed to temporal variations of the reflected fraction of the incident solar wind. In addition, the overshoots of these crossings are enhanced relative to the upstream magnetic field by a substantial factor far greater than the limit predicted by the Rankine-Hugoniot jump conditions, i.e. $B_{max}/B_u \gg 4$ [25]. This has been established to be a typical manifestation of high $M_A$ shocks, underpinning the importance of kinetic over fluid processes, and has not yet been fully explained.

Naturally with single spacecraft observations, we are presented with the challenge of distinguishing between spatial and temporal variability. Repeated shock crossings modulated by the variability in the upstream dynamic pressure (and/or from the downstream region, in this case the internal pressure of a planet's magnetosphere) do occur and some periodic signature is therefore manifested in the data. These however occur at periods that are irregular, markedly larger than those similar to Figure 2 and a downstream "sheath" signature is typically present.

Upon calculating the $M_A$ of all crossings, we have focussed on the highest regime of $M_A \geq 25$ (above the $80^{th}$ percentile) since this has been poorly explored from the observational point of view. Within this regime, we separate the crossings between those that show magnetic field signatures of reformation, given the size and clarity of their quasi-periodic pulses, and those that do not. The criterion is such that there are clear-cut 'reformation cycles' from upstream to downstream of the main transition layer similar to those of Figure 2. Moreover, for each event that exhibits the upstream features similar to Figure 2, we have calculated the period of the cycles (average time between peaks of neighbouring pulses) and find it to be within a narrow range of timescales at which a reformation cycle is predicted to occur. These are shown on Figure 2d as being in the range ~0.2-0.3 $\tau_c$ or ~1.3-1.8 $\Omega_c^{-1}$ (25[th] and 75[th] percentiles respectively), where $\tau_c$ (= $2\pi\Omega_c^{-1}$) is the upstream and undisturbed ion gyroperiod.

Assuming motion in the upstream field, this range (see Figure 2d) is consistent with the proton specular reflected turnaround time for the period of magnetic fluctuations [26]. The shock speeds, where possible, have been determined using a single-spacecraft technique by observing the time it takes for the foot to convect across the spacecraft [27,28]. The shock speeds are listed in Table 1 to show that they are comparable to those of Earth's bow shock. We interpret this periodicity to be associated with reformation which involves a periodic modulation of the reflected ions, with the turnaround time of specularly reflected ions being a natural characteristic period of the foot structure. This is supported by hybrid simulations [29] deducing that at high $M_A$, shock dynamics appear to be likely dominated by ion reflection. Earlier work on hybrid simulations also reports quasi-periodic modulation of the reflected with a period just over $\Omega_c^{-1}$ and suggests for high $M_A$ that ions are reflected dynamically in bunches [30]. More recently, 1-D hybrid simulations running similar plasma parameters to Voyager observations of the Uranian bow shock report localized magnetic field enhancements of a reforming shock occurring at approximately every 1.8 $\Omega_c^{-1}$ [31].

Other possible sources of a periodic signal may originate upstream, e.g. if there were foreshock or solar wind fluctuations. Foreshock driving is not seen and unlikely due to the quasi-perpendicular configuration of the shock and similarly there are no consistent and/or clear signals in the plasma wave instrument that suggests solar wind fluctuation driving. The observations therefore demonstrate that cyclic reformation is the only plausible process and it is probably controlled by specular reflection.

These 54 crossings altogether are highlighted (red or blue) on a parameter space as shown in Figure 3a, and we see a significant fraction of crossings in the population revealing this cyclical feature (red). Crossings that were not highlighted (gray) in this regime (above the $80^{th}$ percentile) were either determined to be quasi-parallel and/or had considerable variability upstream, likely from an active foreshock region and were therefore not included. We have also investigated same-sized populations of the lowest ($M_A < 8$) and middle ($12 \leq M_A < 17$) ranges to find only one and six crossings with such cyclical feature respectively. We focus on the highest $M_A$ regime where the absence or presence of this feature is most clearly distinguishable. Additionally at this regime, we observe multiple upstream peaks over which we can straightforwardly obtain a handle of the periods. The dependence of reformation on $M_A$ is clearly corroborated and this also shows that a high $M_A$ is a necessary but not sufficient condition.

The high-$M_A$ crossings in Figure 3a, however, do not appear to be organized into distinct groups and there is no obvious quantity which separates a reforming from a non-reforming crossing. By taking into account the maximum field $B_{max}$, Figure 3b shows the overshoots $B_{max}/B_u$ versus $M_A$ (of each of the highlighted high-$M_A$ crossings) revealing a correlation, one which has been reported by looking at overshoots of planetary bow shocks with heliocentric distances [32]. The key feature here, on the other hand, is the clear separation between the two groups of crossings and we can infer that shocks undergoing cyclic reformation exhibit a larger overshoot than shocks that do not, for a given $M_A$. This suggests that reflected ions may have a role in amplifying the local magnetic field however the causality between these two processes remains unclear. Nevertheless, the correlations of $B_{max}$ with reformation and the dependence on $M_A \frac{B_u}{B_{max}}$ (and thus $\frac{\sqrt{\rho}u}{B_{max}}$) is explicit.

Indeed, attempting to dissect a single shock crossing here to uncover its microphysics is not likely to be sufficiently instructive; especially in the prevalent analyses with limited particle data sets. We do nonetheless provide a complementary picture of a quasi-perpendicular shock wave's magnetic character from exploiting the rare opportunity of many years' worth of spacecraft data across a large range of $M_A$; up to a regime expected to lie in the same order of magnitude as astrophysical shocks. These observations suggest the prominent role of specular reflection in controlling cyclic reformation. The results also illustrate the dependence of reformation on $M_A$ and the correlation between the magnetic overshoot and $M_A$. The final finding here is the connection between reformation and magnetic field amplification with reforming shocks having a distinctly higher overshoot. This work can complement ongoing theoretical work and simulations. We anticipate our study to provide a deeper insight to collisionless shocks, particularly in astrophysical regimes where they are central to both the structure and dynamics of supernovae.

This work was supported by UK STFC. We acknowledge the support of the Cassini MAG data processing/distribution staff as well as the Royal Astronomical Society.

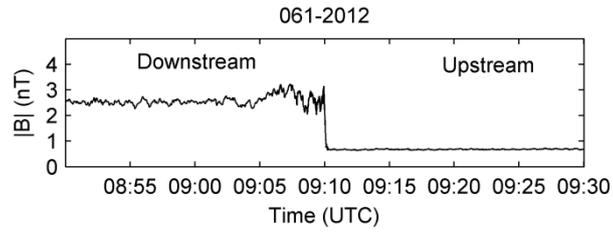

Figure 1 - Magnetic field plot of a Saturnian bow shock crossing. This is a typical example of a quasi-perpendicular shock. Local Time (LT) is 1600 i.e. dusk flank. The plot heading format is "Day of Year (DOY) – Year (YYYY)"

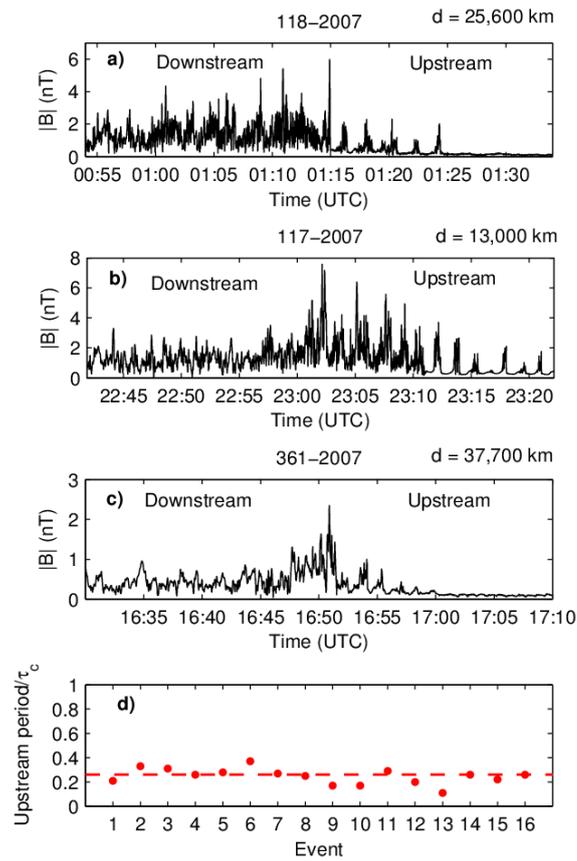

Figure 2 – The top three panels (a-c) are three example magnetic field plots of quasi-perpendicular Saturnian bow shock crossings with reformation cycles upstream. These correspond to events 4, 3 and 13 respectively in panel (d), which displays the average period of the upstream cycles of each event compared to their respective ion gyroperiods. Crossings (a-c) were all situated in the dusk flank at LT 1530, 1530 and 1600 respectively. The foot thickness, *d*, is determined using the formalism from [27].

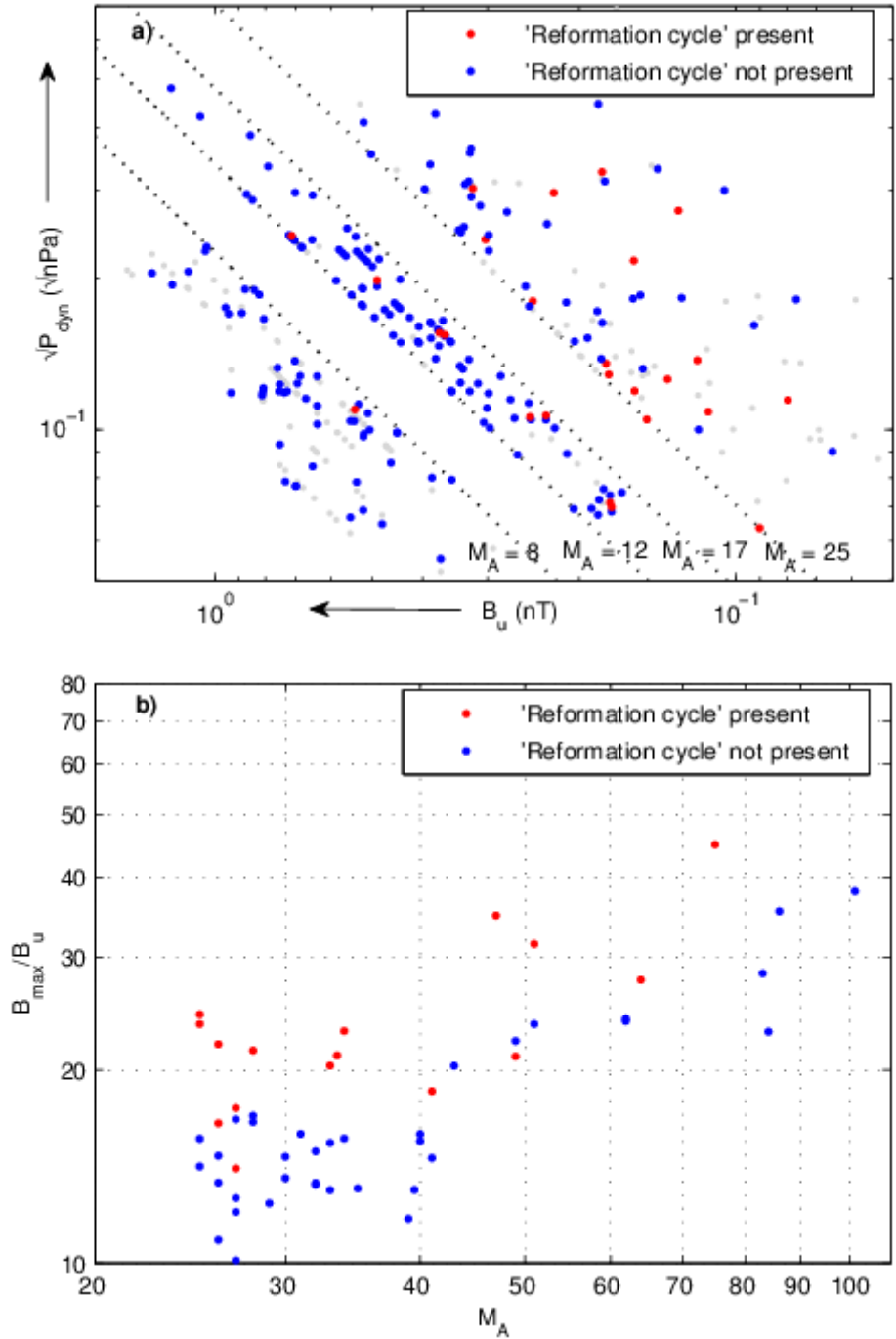

Figure 3 – (a) Parameter space of crossings in three regimes of $M_A$ highlighting in red the events which exhibit the reformation cycles upstream in contrast with the crossings that do not in blue and (b) new parameter space of the crossings in the highest regime, $M_A \geq 25$, with $B_{max}$.

| Event | Time (YYYY-DOY-HHMM) | $\theta_{Bn}$ | $n_p{}^{a,c}$ (cm$^{-3}$) | $V_n{}^{b,c}$ (km/s) | $V_{shock}$ (km/s) | $M_A{}^d$ | Upstream period/$\tau_c$ | Upstream period/$\Omega_c{}^{-1}$ |
|---|---|---|---|---|---|---|---|---|
| 1 | 2007-071-0608 | 79 | 0.23 | 376 | -3 | 27 | 0.21 | 1.31 |
| 2 | 2007-117-1911 | 88 | 0.27 | 489 | - 6 | 64 | 0.33 | 2.05 |
| 3 | 2007-117-2302 | 79 | … | 400 | -9 | 47 | 0.31 | 1.94 |
| 4 | 2007-118-0114 | 61 | 0.28 | 401 | -7 | 74 | 0.26 | 1.65 |
| 5 | 2007-152-1837 | 55 | … | 424 | +55 | 41 | 0.28 | 1.73 |
| 6 | 2007-153-1505 | 85 | … | … | … | 26 | 0.37 | 2.34 |
| 7 | 2007-154-0800 | 80 | 0.05 | 374 | -6 | 25 | 0.27 | 1.69 |
| 8 | 2007-155-0046 | 83 | … | … | … | 33 | 0.25 | 1.56 |
| 9 | 2007-265-2131 | 89 | 0.02 | 361 | -13 | 25 | 0.17 | 1.07 |
| 10 | 2007-340-0445 | 77 | … | 433 | -3 | 49 | 0.17 | 1.05 |
| 11 | 2007-341-1251 | 54 | … | … | … | 27 | 0.29 | 1.83 |
| 12 | 2007-360-2157 | 78 | … | … | … | 34 | 0.20 | 1.24 |
| 13 | 2007-361-1650 | 86 | … | 418 | +12 | 51 | 0.11 | 0.69 |
| 14 | 2007-362-0459 | 90 | … | 410 | -5 | 27 | 0.26 | 1.66 |
| 15 | 2008-046-0006 | 78 | 0.05 | … | … | 26 | 0.22 | 1.39 |
| 16 | 2008-096-2011 | 67 | 0.12 | … | … | 34 | 0.26 | 1.62 |
| | | | | | | **Median** | 0.26 | 1.64 |
| | | | | | | **STD** | 0.066 | 0.416 |

[a] Unshocked proton density.

[b] $V_n \equiv \mathbf{V} \cdot \hat{\mathbf{n}}$ (shock rest frame).

[c] Entries with a hyphen indicate where observations are unavailable or ambiguous.

[d] Calculated where both $n_p$ and $V_n$ measurements are available; otherwise derived from the shock model [23].

Table 1 – Reforming shock events